\documentstyle[aps,prl,epsf,twocolumn,floats]{revtex}
\begin{document}
\draft
\twocolumn[\csname @twocolumnfalse\endcsname
\widetext

\title{
Critical local-moment fluctuations, anomalous exponents,
and $\omega/T$ scaling in the Kondo problem with a pseudogap}
\author{Kevin Ingersent$^{(a)}$ and Qimiao Si$^{(b)}$}
\address{$^{(a)}$Department of Physics, University of Florida, Gainesville,
FL 32611--8440 \\
$^{(b)}$Department of Physics \& Astronomy, Rice University, Houston,
TX 77005--1892}

\maketitle

\begin{abstract}
Experiments in heavy-fermion metals and related theoretical work
suggest that critical local-moment fluctuations can play an important
role near a zero-temperature phase transition.
We study such fluctuations at the quantum critical point of a Kondo impurity
model in which the density of band states
vanishes as $|\epsilon|^r$ at the Fermi energy ($\epsilon=0$).
The local spin response is described by a set of critical exponents
that vary continuously with $r$.
For $0<r<1$, the dynamical susceptibility exhibits $\omega/T$ scaling
with a fractional exponent, implying that the critical point is interacting.
\end{abstract}
]

\narrowtext

A number of stoichiometric (or nearly stoichiometric) heavy-fermion metals
exhibit non-Fermi-liquid behavior when tuned to the vicinity of
a magnetic quantum critical point (QCP)
\cite{Mathur,vonLohneysen,Steglich,Stewart,Schroder,Stockert}.
An important clue as to the nature of the quantum criticality
has come from neutron scattering experiments\cite{Schroder,Stockert}
near the magnetic QCP of CeCu$_{6-x}$Au$_x$.
At the (rather small) critical Au concentration, $x_c \approx 0.1$, the
dynamical spin susceptibility is highly unusual in two respects:
First, it satisfies $\omega/T$ scaling.
Second, the frequency and temperature dependence obeys a fractional power law,
described by the same anomalous exponent over essentially the entire Brillouin
zone.
There are indications that the stoichiometric system ${\rm YbRh_2Si_2}$
behaves similarly\cite{Steglich}.
These experiments directly
suggest\cite{Schroder,Stockert,Coleman,SiSmithIngersent,UPdCu}
that the fluctuations of the individual local moments are also critical.
While the standard Kondo behavior of local moments in simple metals has been
studied extensively over the past four decades and is well
understood \cite{Hewson}, the physics of {\it critical local-moment
fluctuations} is largely unexplored.
It is therefore highly desirable to identify
models that are amenable to controlled theoretical study.

Further motivation for studying critical local-moment fluctuations
comes from related theoretical work.
We have recently shown\cite{lcp} that competition between the
Kondo and Ruderman-Kittel-Kasuya-Yosida interactions in a Kondo lattice model
generates a new class of QCP, which we argue
explains the aforementioned experiments in heavy fermions.
Here, not only are the long-wavelength spin fluctuations critical but
so also are the local-moment fluctuations; the weight of the
Kondo resonance goes to zero at the QCP.
The existence of critical local fluctuations distinguishes such a ``locally
critical point'' from the standard picture based on a spin-density-wave
transition\cite{Hertz,Millis}.
To fully elucidate the properties of the locally critical point,
one must construct a Ginzburg-Landau description.
This requires one to understand the precise nature of the critical local
mode that characterizes the destruction of the Kondo effect.
The first step towards this goal is to develop an intuition about critical
local-moment fluctuations in simpler models.

This paper addresses just such a model: the single-impurity Kondo problem
with a power-law pseudogap.
The model has a quantum phase transition at a finite Kondo coupling
\cite{Withoff}.
We
show that the QCP exhibits critical local-moment fluctuations and an
associated destruction of the Kondo effect very similar to those present
at the locally critical point of the Kondo lattice; the local susceptibility
displays $\omega/T$ scaling with a fractional exponent.
The many-body spectrum of this model can be calculated exactly, an important
virtue which should significantly aid the identification of the critical local
mode.

The Kondo model for a single spin-$\frac{1}{2}$ impurity
coupled to a conduction band is described by the Hamiltonian
\begin{eqnarray}
{\cal H}_{\text{K}} =
&& \sum_{{\bf k},\sigma} \epsilon_{\bf k}
    c_{{\bf k}\sigma}^{\dagger} c_{{\bf k}\sigma}
    + \frac{J}{2} {\bf S} \cdot \sum_{\sigma,\sigma'} c_{0\sigma}^{\dagger}
    \bbox{\tau}_{\sigma \sigma'} c_{0\sigma'} \nonumber \\[-1ex]
&& + \; V \sum_{\sigma} c_{0\sigma}^{\dagger} c_{0\sigma},
\label{Kondo_Hamiltonian}
\end{eqnarray}
where $\bf S$ is the impurity spin operator, $c^{\dagger}_{0\sigma}$ creates
an electron with spin $z$ component $\sigma$
($= \pm\frac{1}{2}$) at the impurity site,
and $\tau^i_{\sigma \sigma'}$ ($i=x, y, z$)
is the standard Pauli matrix.
$J$ is the Kondo exchange coupling, while $V$ parametrizes nonmagnetic
potential scattering from the impurity site.

In the power-law version of this model, the conduction band is described by
the (oversimplified) particle-hole-symmetric density of states \cite{energies}
\begin{eqnarray}
\rho(\epsilon) = \left\{
   \begin{array}{ll}
       \rho_0 |\epsilon|^r \quad & \mbox{for } |\epsilon| \le 1, \\[0.5ex]
       0 & \mbox{for } |\epsilon| > 1.
   \end{array} \right.
\label{dos}
\end{eqnarray}
In a metal ($r=0$), any antiferromagnetic Kondo coupling $J > 0$ causes
the impurity moment to be quenched at temperature $T=0$ \cite{Hewson}.
With a pseudogap ($r>0$), by contrast, quenching occurs only for
$J > J_c > 0$ \cite{Withoff}.
The strong-coupling (i.e., $J>J_c$) and weak-coupling ($J<J_c$) properties
of the power-law Kondo model have been studied extensively
\cite{Withoff,Cassanello,Chen,Gonzalez-Buxton,Bulla,VojtaBulla}.
Here, instead, we study the critical behavior at
$J=J_c(r,V)$ using
numerical calculations and controlled analytical approximations.
(The model appears not to be integrable \cite{Andrei};
it also lacks conformal invariance.)
This paper supersedes an earlier preprint\cite{cond-mat},
which discussed only static critical properties, and focused on the
large-$N$ limit of Eq.~(\ref{Kondo_Hamiltonian}).
More recent work\cite{Vojta} has addressed
QCPs in multi-channel Kondo problems with a pseudogap.

\begin{figure}
\centerline{\epsfxsize=55mm\epsfbox{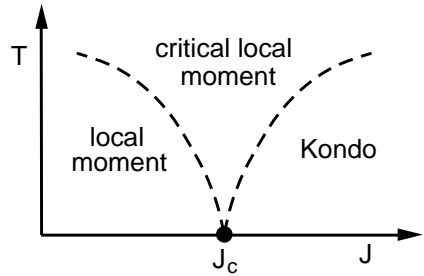}}
\vspace{1ex}
\caption{%
Schematic phase diagram showing the vicinity of the quantum critical point
of the pseudogap Kondo model.
}
\label{fig:phase}
\end{figure}

Under conditions of strict particle-hole symmetry
[$V=0$ in Eq.~(\ref{Kondo_Hamiltonian})], a symmetric critical point
(SCP) separates the weak- and strong-coupling regimes for all
$0\!<r\!<\frac{1}{2}$; the strong-coupling regime and the SCP
both vanish for $r\ge\frac{1}{2}$ \cite{Chen}.
The SCP is also encountered away from particle-hole symmetry for
$0<r\le r^* \approx 0.375$; for $r > r^*$, however, there is an
asymmetric critical point (ACP) distinct from the SCP \cite{Gonzalez-Buxton}.
In all cases, the transition can be schematically represented
as shown in Fig.~\ref{fig:phase}.

{\em Local vs impurity susceptibility.}---Our analysis begins with the
observation that the quantum critical behavior reveals itself, not in the
response to a uniform magnetic field~$H$, but rather in that to a local
magnetic field~$h$ coupled solely to the impurity \cite{H_mag,units}.
These responses are measured, respectively, by the static
impurity susceptibility
$\chi_{\text{imp}}\!=\!-\partial^2 F_{\text{imp}}/\partial H^2|_{H=h=0}$,
and the static
local susceptibility
$\chi_{\text{loc}}\!=\!-\partial^2 F_{\text{imp}}/\partial h^2|_{H=h=0}$,
where $F_{\text{imp}}$ is the impurity contribution to the free energy.
Numerical renormalization-group (NRG) results \cite{Chen,Gonzalez-Buxton}
indicate that, whereas $\lim_{T\rightarrow 0}T\chi_{\text{imp}}$ undergoes
a jump as $J$ passes through $J_c$, $\lim_{T\rightarrow 0}T\chi_{\text{loc}}$
goes continuously to zero as the critical coupling is approached from below,
and $\lim_{T\rightarrow 0}T\chi_{\text{loc}}\!=\!0$ for all $J>J_c$.
(The same distinction
also holds
in the large-$N$ limit \cite{cond-mat}.)

{\em Static critical properties.}---Given that the local field~$h$ (rather
than the uniform field~$H$) acts as a scaling variable, we define exponents
$\beta$, $\gamma$, $\delta$, and $x$, describing the critical behavior of the
the local susceptibility and the local-moment amplitude $M_{\text{loc}}=
\langle S_z\rangle\!=\!-\partial F_{\text{imp}}/\partial h|_{H=0}$:
\begin{eqnarray}
M_{\text{loc}}(J < J_c,T=0,h=0)
&\propto& (J_c-J)^{\beta}, \nonumber\\
\chi_{\text{loc}}(J > J_c,T=0) &\propto& (J-J_c)^{-\gamma},
\nonumber\\[-1.75ex]
\label{exponents} \\[-1.75ex]
M_{\text{loc}}(J=J_c,T=0) &\propto& | h |^{1/\delta}, \nonumber\\
\chi_{\text{loc}}(J=J_c) &=&
C_{\text{static}} T^{-x}. \nonumber
\end{eqnarray}

Using a generalization \cite{Gonzalez-Buxton} of Wilson's NRG method
\cite{Wilson} to treat the density of states in Eq.~(\ref{dos}),
we have computed $M_{\text{loc}}$ and $\chi_{\text{loc}}$
for $r$ between $0.1$ and $2$.
For $0<r<1$, we find that $M_{\text{loc}}$ and $\chi_{\text{loc}}$ obey
Eqs.~(\ref{exponents}) near both the SCP and the ACP,
establishing the continuous nature of these phase transitions \cite{Withoff}.
For $r>1$, by contrast, $M_{\text{loc}}$ undergoes a jump at the transition.

\begin{table}[b]
\begin{tabular}{ll@{\hspace*{-1em}}dlll}
$~r$ &
$~~~\beta$ &
$\gamma$ &
\multicolumn{1}{c}{$1/\delta$} &
\multicolumn{1}{c}{$x$} &
\multicolumn{1}{c}{$R$}
\\[0.5ex]
\hline
0.1  &           & 10.63(2) & 0.00565 & 0.9888 & 0.040(3) \\
0.15 & 0.1033(2) & 7.476(2) & 0.01367 & 0.9730 \\
0.2  & 0.1600(2) & 5.899    & 0.02645 & 0.9485 & 0.17 \\
0.3  & 0.3548(2) & 4.441    & 0.0740  & 0.8622 \\
0.4  & 0.9140    & 4.018(3) & 0.1852  & 0.6875 & 0.54 \\
0.45 & 1.982(5)  & 4.335(3) & 0.3134  & 0.5228
\end{tabular}
\vspace{2ex}
\caption{%
Properties of the symmetric critical point, obtained from NRG calculations.
See the text for definitions of the exponents $\beta$, $\gamma$, $\delta$,
and $x$ (all determined using a band-discretization $\Lambda=9$),
and of the ratio $R$ (calculated for $\Lambda=3$).
Parentheses surround the estimated nonsystematic error in the last digit
(equal to 1 where omitted).
}
\label{tab:symm}
\end{table}

\begin{figure}
\centerline{\epsfxsize=73mm\epsfbox{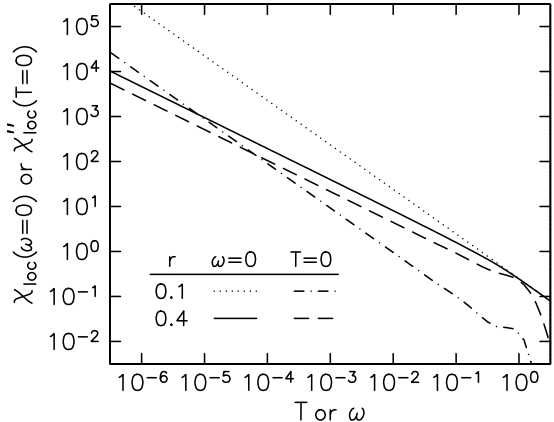}}
\vspace{1ex}
\caption{%
$\chi_{\text{loc}}
(\omega\!=\!0)$ vs $T$ and
$\chi_{\text{loc}}''(T\!=\!0)$ vs $\omega$ at the sym\-metric critical point:
NRG results for $\Lambda=3$, $r=0.1$ and $0.4$ \protect\cite{units}.
For a given $r$,
$\chi_{\text{loc}}(\omega=0,T\ll 1)$
and
$\chi_{\text{loc}}''(|\omega|\ll 1,T=0)$
are described by equal exponents, $x=y$.
}
\label{fig:chi_loc}
\end{figure}

Table~\ref{tab:symm} lists exponents for the SCP, along with their estimated
nonsystematic (numerical-rounding and slope-fitting) errors.
Data at $J=J_c$ exhibit power laws over at least five decades of $h$ and $T$
(e.g., see $\chi_{\text{loc}}$ vs $T$ in Fig.~\ref{fig:chi_loc}),
allowing precise determination of $\delta$ and $x$.
The uncertainty in $\beta$ and $\gamma$ is greater because rounding error
cuts off the power laws as $J$ approaches $J_c$.
Most runs were performed for an NRG discretization parameter $\Lambda=9$,
retaining all states within an energy $50T$ of the ground state \cite{units}.
To estimate the systematic discretization errors, a few runs were performed
using a value $\Lambda=3$ lying closer to the continuum limit
($\Lambda=1$) but requiring much more computer time.
Critical exponents computed for $\Lambda=3$ and $\Lambda=9$ only narrowly
fail to agree within their estimated nonsystematic errors, so we believe
that the $\Lambda=9$ exponents approximate the continuum values quite well.
Restricting the power-law form of $\rho(\epsilon)$ to a finite
region around the Fermi energy to better approximate real systems does
not alter the critical exponents.

The exponents listed in Table~\ref{tab:symm} have nontrivial $r$ dependence.
To better understand these exponents, we show that they satisfy certain
hyperscaling relations, which can be derived in a standard fashion.
We expect the singular component of the free energy to take the
form \cite{units}
\begin{eqnarray}
F_{\text{imp}}
= T f\left(|J\!-\!J_c|/T^a, \, |h|/T^b\right).
\label{F_s}
\end{eqnarray}
Using Eq.~(\ref{F_s}), one readily finds that
$\beta = (1-b)/a$,
$\gamma = (2b-1)/a$,
$\delta = b/(1-b)$,
and $x = 2 b - 1$.
These expressions lead to a pair of hyperscaling
relations among the critical exponents, e.g.,
\begin{equation}
\delta = (1 + x) / (1 - x), \quad
\beta = \gamma (1 - x) / (2 x).
\label{hyperscaling}
\end{equation}
In all cases, the exponents listed in Table~\ref{tab:symm} satisfy the
hyperscaling relations to the accuracy of our calculations.

\begin{table}[t]
\begin{tabular}{@{\hspace*{1em}}llllll}
$~r$ &
\multicolumn{1}{c}{$\beta$} &
\multicolumn{1}{c}{$\gamma$} &
\multicolumn{1}{c}{$1/\delta$} &
\multicolumn{1}{c}{$x$} &
\\[0.5ex]
\hline
0.4  & 0.58     & 3.12     & 0.1570(2) & 0.7285(5) \\
0.6  & 0.188    & 1.41     & 0.1168    & 0.7905(5) \\
0.8  & 0.077(2) & 1.108(4) & 0.0645(7) & 0.8795(5) \\
0.9  & 0.039(2) & 1.025(3) & 0.035     & 0.928(2)
\end{tabular}
\vspace{2ex}
\caption{%
Exponents at the asymmetric critical point,
from NRG calculations using a discretization parameter $\Lambda=9$.
The symbols are explained in Table~\protect\ref{tab:symm}.
}
\label{tab:asymm}
\end{table}

Table~\ref{tab:asymm} lists critical exponents at the ACP.
Comparison with Table~\ref{tab:symm} shows that within their range of
coexistence ($0.375\alt r< \frac{1}{2}$), the SCP and ACP have different
exponents.
For all $r<1$, the hyperscaling relations, Eqs.~(\ref{hyperscaling}), are
obeyed to within estimated errors.
It proves difficult to determine the critical behavior for $r=1$, where
there are logarithmic corrections to scaling \cite{Cassanello}.
For $1<r<2$, $M_{\text{loc}}$ is no longer critical, but $\chi_{\text{loc}}$
is described by exponents $\gamma=2-r$ and $x=1$,
values that are consistent with Eq.~(\ref{F_s}) if $a=1/(2-r)$ and $b=1$.

{\em Dynamical critical properties.}---We have also computed the imaginary
part of the dynamical local susceptibility,
$\chi_{\text{loc}}''(J\!=\!J_c,\omega,T)$.
Figure~\ref{fig:chi_loc} shows some of our zero-temperature results at the SCP.
The low-frequency NRG data at both the SCP and the ACP fit the form
\begin{equation}
\chi_{\text{loc}}''(J\!=\!J_c,\omega,T\!=\!0) =
C_{\text{dynamic}}
|\omega|^{-y} \, \text{sgn} \, \omega.
\label{exponent-y}
\end{equation}
For $0\!<\!r\!<\!1$, we find $y=x$ to within numerical error,
an equality that is consistent with a scaling form:
$\chi_{\text{loc}}(J\!=\!J_c,\omega,T) = T^{-x} X(\omega/T)$.
Such an $\omega/T$ scaling cannot hold for $1\!<\!r\!<\!2$,
where we find $y = \gamma < x$.

\begin{figure}
\centerline{\epsfxsize=82mm\epsfbox{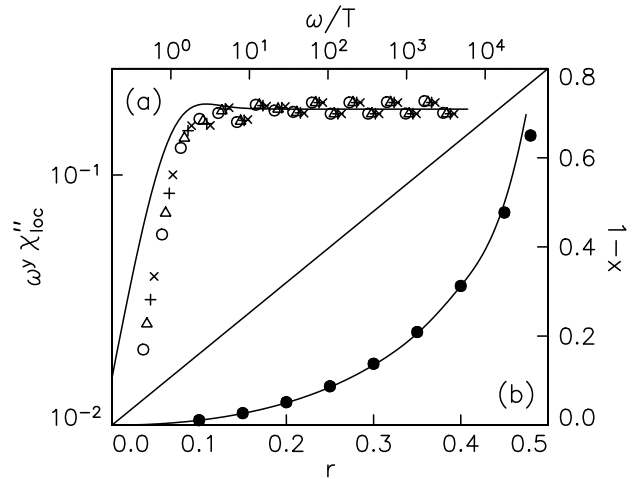}}
\vspace{1ex}
\caption{%
Comparison between NRG and small-$r$ results for the symmetric critical point.
(a) $\omega^y \chi_{\text{loc}}''(J\!=\!J_c)$ vs $\omega/T$
\protect\cite{units}:
NRG results for $r=0.4$ and $T/D = 10^{-7}$, $10^{-6}$, $10^{-5}$,
and $10^{-4}$ (symbols) and the prediction of
Eq.~(\protect\ref{chi-omega-T}) (solid line).
(b) Exponent $1-x$ vs $r$: NRG data from Table~\protect\ref{tab:symm}
(symbols) and the prediction of Eq.~\protect(\ref{x:small_r}), in the form
of a spline fit (solid line) through numerical values for $(\rho_0 J_c)^2$,
extrapolated to the continuum limit
$\Lambda=1$ (see \protect\onlinecite{Gonzalez-Buxton}).
}
\label{fig:NRG_vs_small-r}
\end{figure}

For a given $r$ between 0 and 1, the data for
$\chi_{\text{loc}}''(J\!=\!J_c,\omega,T)$ collapse onto a single function
of $\omega/T$, as illustrated in Fig.~\ref{fig:NRG_vs_small-r}(a).
This does not conclusively establish $\omega/T$ scaling because the
NRG method is unreliable in the regime $|\omega|\lesssim T$.
For small $r$, however, we can use a different approach to confirm the
existence of scaling.

For small $r$, the local spin-spin correlation function can be calculated
algebraically by a procedure analogous to the standard $\epsilon$
expansion \cite{Wilson-Kogut}: Since the critical coupling is small, the fixed
point is accessible via an expansion in $\rho_0 J_c \approx r$.
The unperturbed reference point ($\rho_0 J_c=r=0$)
describes the standard Kondo problem, so
the perturbation series for $\chi_{\rm loc}$ at the
critical point contains logarithmic singularities.
To leading logarithmic order, we find
$\chi_{\rm loc} (\tau) = {1 \over 4}
[ 1 - (\rho_0 J_c)^2 \ln (\pi T \tau_0 /
\sin \pi T \tau ) ]$, where $\tau_0 \approx \rho_0$.
This leads to
\begin{equation}
\chi_{\text{loc}}
(\tau)
= {1 \over 4}
\left( { {\pi T \tau_0} \over {\sin {\pi T \tau}}}
\right )^{\eta} , \quad \eta = (\rho_0 J_c)^2,
\label{chi-tau:small_r}
\end{equation}
and to a dynamical spin susceptibility
having the asymptotic low-energy, low-frequency form
\begin{eqnarray}
\chi_{\text{loc}}
(\omega, T) =
{\tau_0^{\eta} \sin (\pi \eta / 2 ) \over 2 (2\pi T)^{1-\eta}}
\, B\left({\eta \over 2}-i {\omega \over {2 \pi T}}, 1-\eta \right) ,
\label{chi-omega-T}
\end{eqnarray}
$B$ being the Euler beta function.

It follows from Eq. (\ref{chi-omega-T})
that, for small $r$, the static local susceptibility
and the imaginary part of the local susceptibility at
$T=0$ have the forms specified by Eqs.~(\ref{exponents})
and~(\ref{exponent-y}), respectively, with exponents
\begin{equation}
x = y = 1 - \eta = 1 - (\rho_0 J_c)^2,
\label{x:small_r}
\end{equation}
and a universal (cutoff-independent)
amplitude ratio
\begin{equation}
  R \equiv \frac{C_{\text{dynamic}}}{C_{\text{static}}} =
    \frac{\pi^{2-\eta}\,\Gamma(1\!-\!\eta/2)}{2^\eta \sin(\pi \eta/2)
    \Gamma(1\!-\!\eta) \Gamma(\eta) \Gamma(\eta/2)}
.
\label{R:small_r}
\end{equation}

These results can be compared with our NRG data:
(i) As mentioned above, we find that $y=x$ is obeyed to
the accuracy of our calculations for all $0<r<1$.
(ii) The calculated $\chi_{\text{loc}}$ obeys Eq.~(\ref{chi-omega-T})
within the range $\omega \gg T$ where the NRG method is reliable
[see Fig.~(3(a)].
(iii) The numerical values of $x$ agree remarkably well with the small-$r$
result $x=1-(\rho_0 J_c)^2$, even when $r$ (and, hence, $\rho_0 J_c$) is not
small [see Fig.~3(b)].
(iv) The $R$ values in Table~\ref{tab:symm} fit Eq.~(\ref{R:small_r}) to
within 25\%, a reasonable level of agreement given that the systematic errors
in prefactors and critical couplings computed using the NRG are generally
greater than the errors in critical exponents.

One of our key conclusions is that the dynamical spin susceptibility
satisfies $\omega/T$ scaling, as shown by Eq.~(\ref{chi-omega-T})
for small $r$, and supported by the equality of the exponents $x$ and $y$
for all $0<r<1$.
This has important implications for the field-theoretical
description of the QCP. It indicates that a suitably defined
relaxation rate is linear in temperature, which can only come about if
the Ginzburg-Landau action, written in terms of the critical local
modes, contains nonlinear couplings that are relevant in the
renormalization-group sense\cite{Sachdev-book}.
The fixed point must be interacting.
For $r>1$, the relaxation rate ($\propto T^{x/y}$) is superlinear,
which is consistent with a Gaussian fixed point.

Finally, we note that the $r=1$ pseudogap Kondo problem is also important
for impurities in high-$T_c$ superconductors \cite{STM-Kondo,VojtaBulla}.
The quantum critical regime will likely become accessible via scanning
tunneling microscopy once measurements are extended to higher temperatures.

In summary, we have combined numerical and analytical approaches to obtain
a consistent picture of the critical properties of the Kondo problem with a
conduction-electron density of states proportional to $|\epsilon|^r$.
At the QCP, the weight of the Kondo resonance has just gone to zero;
as a result, the local-moment fluctuations are critical.
In addition, the dynamical spin susceptibility at the QCP displays
$\omega/T$ scaling with an anomalous exponent for all $0<r<1$.
These features parallel those of the locally critical point in the Kondo
lattice \cite{lcp}.
Thus, the pseudogap Kondo model provides a testing ground for studying
critical local-moment fluctuations of the type seen in certain
heavy-fermion metals\cite{Steglich,Stewart,Schroder,Stockert}.
The exact many-body spectrum of this impurity model
can be calculated using NRG techniques,
which should prove particularly useful in identifying the proper
local modes for characterizing the QCP. This, in turn, should
shed much new light on the Ginzburg-Landau
description of the locally critical point of the Kondo lattice.
These important issues are left for future work.

We would like to thank C.\ R.\ Cassanello,
E.\ Fradkin, and J.\ W.\ Wilkins for useful discussions.
This work has been supported in part by NSF Grant No.\ DMR-9316587 (K.I.),
and by the Robert A. Welch Foundation, NSF Grant No.\ DMR-0090071,
Research Corporation, and TCSUH (Q.S.).
Q.S.\ also acknowledges the hospitality of Argonne National Laboratory, the
University of Chicago, and the University of Illinois at Urbana-Champaign.


\begin{references}
\bibitem{energies}
    All energies are expressed in units of the half-bandwidth and measured
    from the Fermi energy, $\epsilon = 0$.
\bibitem{units}
    We employ units in which $\hbar=1$, $k_B=1$, and $g\mu_B=1$.
\end{references}
\end{document}